\documentstyle[12pt,epsfig]{article}
\newlength{\dinwidth}
\newlength{\dinmargin}
\newcommand{\ba}{\begin{array}}
\newcommand{\ea}{\end{array}}
\newcommand{\beq}{\begin{equation}}
\newcommand{\eeq}{\end{equation}}
\newcommand{\bea}{\begin{eqnarray}}
\newcommand{\eea}{\end{eqnarray}}

\def\S{{\bf S}}

\def\bce{\begin{center}}
\def\ece{\end{center}}

\def\nonu{\nonumber}

\def\pa{\partial}
\def\al{\alpha}
\def\be{\beta}

\def\De{\Delta}

\def\th{\theta}

\def\ph{\phi}

\def\ps{\psi}

\def\R{{\bf R}}

\def\S{{\bf S}}

\begin{document}
\thispagestyle{empty}
\addtocounter{page}{-1}
\begin{flushright}
{\tt hep-th/0205008}\\
\end{flushright}
\vspace*{1.3cm}
\centerline{\Large \bf More on Penrose Limit of
$AdS_4 \times Q^{1,1,1}$}
\vspace*{1.5cm} 
\centerline{ \bf Changhyun Ahn
}
\vspace*{1.0cm}
\centerline{\it 
Department of Physics,}
\vskip0.3cm 
\centerline{  \it Kyungpook National University,}
\vskip0.3cm
\centerline{  \it  Taegu 702-701, Korea }
\vspace*{0.3cm}
\centerline{\tt ahn@knu.ac.kr 
}
\vskip2cm
\centerline{\bf  abstract}
\vspace*{0.5cm}

We consider a Penrose limit of $AdS_4 \times Q^{1,1,1}$
that provides the pp-wave geometry equal to the one in the Penrose limit
of $AdS_4 \times \S^7$. 
We expect that there exists a subsector of 
three dimensional ${\cal N}=2$ dual gauge theory which 
has enhanced ${\cal N}=8$ maximal supersymmetry.
We identify operators in the ${\cal N}=2$ gauge theory with
11-dimensional supergravity KK excitations in the pp-wave geometry and 
describe how both the chiral multiplets and
semi-conserved multiplets fall into 
${\cal N}=8$ supermultiplets.

\vspace*{4.0cm}

\baselineskip=18pt
\newpage

\section{Introduction}

Recently, it was found in \cite{bmn} 
that the large $N$ limit of a subsector of
four dimensional ${\cal N}=4$ $SU(N)$ supersymmetric gauge theory is dual to
type IIB string theory in the pp-wave background \cite{bfhpetal,bfhpetal1}.
In the $N \rightarrow \infty$ limit with the finiteness of string coupling
constant, this subspace of the gauge theory and the operator algebra are
described by string theory in the pp-wave geometry.
By considering a scale limit of the geometry near a null geodesic in $AdS_5 
\times \S^5$, it leads to a truncation to 
the appropriate subspace of the gauge 
theory. The operators in the subsector of ${\cal N}=4$ gauge theory can be 
identified with the excited string states in the pp-wave background. 
There are various and relevant papers \cite{mt}-\cite{tt1} 
related to the work of \cite{bmn}.

There exist some results \cite{ikm,go,zs}
on the Penrose limit of $AdS_5 \times T^{1,1}$ that
gives the pp-wave geometry of $AdS_5 \times \S^5$. 
There is a subsector of ${\cal N}=1$ gauge theory
that contains an enhanced ${\cal N}=4$ supersymmetry. The corresponding 
operators in the gauge theory side were identified with the stringy 
excitations in the pp-wave geometry and some of the gauge theory operators
are combined into ${\cal N}=4$ supersymmetry multiplets \cite{go}.  

In this paper, we consider a similar duality
that is present between a certain three dimensional ${\cal N}=2$
gauge theory and 11-dimensional supergravity theory
in a pp-wave background with the same spirit as in \cite{ikm,go,zs}. 
We describe this duality by taking a scaling limit of both
11-dimensional supergravity on $AdS_4 \times Q^{1,1,1}$(
$Q^{1,1,1}$ was found in \cite{dfv} first to show an Einstein metric)
and three dimensional superconformal field theory.
Three-dimensional theory  consists of an ${\cal N}=2$ 
$SU(N)_1 \times SU(N)_2 \times SU(N)_3$ gauge theories
with three kinds of chiral fields $A_i, i=1,2$ transforming in the 
$\bf (N, \overline{N},1)$ representation, $B_j, j=1,2$ transforming in the 
$\bf (1,N, \overline{N})$ representation and $C_k, k=1,2$
transforming in the $\bf (\overline{N},1,N)$ representation.
This gives the theory of \cite{fabbrietal} that lives on $N$ M2-branes 
at the conical singularity of a Calabi-Yau four-fold. 
The complete analysis on the spectrum of
$AdS_4 \times Q^{1,1,1}$ was studied by Merlatti \cite{mer}. 
The scaling limit is obtained by considering the geometry near a null 
geodesic carrying large angular momentum in the $U(1)_R$
isometry of the $Q^{1,1,1}$ space which is dual to the $U(1)_R$ R-symmetry
in the ${\cal N}=2$ superconformal field theory.
In section 2, we consider and review the scaling limit around
a null geodesic in $AdS_4 \times Q^{1,1,1}$ and obtain an ${\cal N}=8$ 
maximally supersymmetric pp-wave 
background of $AdS_4 \times \S^7$ described by \cite{gns}.
In section 3, we identify supergravity KK excitations obtained by
Merlatti \cite{mer} in the Penrose limit
with gauge theory operators.
In section 4, we summarize our results.

\section{Penrose Limit of $AdS_4 \times Q^{1,1,1}$}

Let us start with the supergravity solution dual to
the ${\cal N}=2$ superconformal field theory \cite{fabbrietal}.
By putting a large number of $N$ coincident M2-branes at the
conifold singularity and taking the near horizon limit,
the metric becomes  that \cite{ffgt,fabbrietal} of 
$AdS_4 \times Q^{1,1,1}$(See also \cite{np,pp,pp1})
\bea
ds_{11}^2 =ds_{AdS_4}^2 +ds_{Q^{1,1,1}}^2,
\label{metric}
\eea
where
\bea
ds_{AdS_4}^2 & = & L^2 \left( -\cosh^2 \rho \; d t^2 +d \rho^2 + 
\sinh^2 \rho \; d \Omega_2^2 \right),
\nonu \\
ds_{Q^{1,1,1}}^2 & = & 
\frac{L^2}{16}  \left( d\ps +\sum_{i=1}^{3}  \cos \th_i d \ph_i \right)^2
 + \frac{L^2}{8}  \sum_{i=1}^{3}  \left( d \th_i^2+\sin^2 \th_i d \ph_i^2
\right),
\nonu
\eea
where $ d \Omega_2$ is the volume form of a unit two sphere $\S^2$
and the curvature radius\footnote{Comparing with the one in \cite{ahnplb},
the $L^6$ here is related to $3L/8$ there. Note that the ratio of
the volume of $\S^7$ to the volume of $Q^{1,1,1}$ is $8/3$. The overall
factor $(8/3)^{1/3}$ is absorbed in the 11-dimensional metric $ds_{11}^2$.} 
$L$ of $AdS_4$ is given in terms of the number of M2-branes by
$(2L)^6= 32 \pi^2 \ell_p^6 N$. Topologically $Q^{1,1,1}$ is a $U(1)$ bundle
over $\S^2_1 \times \S^2_2 \times \S^2_3$. The base is parametrized by
the spherical coordinates $(\theta_i, \phi_i), i=1,2,3$ to parametrize
$i$-th two sphere, as usual, and the angle $\psi$ which has period $4\pi$
parametrizes the $U(1)$ Hopf 
fiber.
The $SU(2)_1 \times SU(2)_2 \times SU(2)_3 \times U(1)_R$ isometry
 group of $Q^{1,1,1}$ corresponds to  $SU(2)_1 \times SU(2)_2 
\times SU(2)_3 $ global symmetry and $U(1)_R$ $R$-symmetry of the dual 
superconformal field theory of \cite{fabbrietal}.

Let us make a scaling limit around a null geodesic in 
 $AdS_4 \times Q^{1,1,1}$ that rotates along the $\psi$ coordinate
of $Q^{1,1,1}$ whose shift symmetry corresponds to
the $U(1)_R$ symmetry in the dual superconformal field theory.  
Similar analysis in $AdS_5 \times T^{1,1}$ appeared in \cite{ikm,go,zs}.
Let us introduce
coordinates which label the geodesic   
\bea
x^{+}  =  \frac{1}{2} \left( t +\frac{1}{4} \left( \psi + \phi_1
+\phi_2 +\phi_3\right) \right), \qquad
x^{-}  =   \frac{L^2}{2} \left( t -\frac{1}{4} \left( \psi + \phi_1
+\phi_2 +\phi_3\right) \right),
\nonu
\eea
and make a scaling limit around $\rho=0=\theta_1=\theta_2=\theta_3$
in the above geometry (\ref{metric}).
By taking the limit $L \rightarrow \infty$ while 
rescaling the coordinates 
\bea
\rho=\frac{r}{L}, \qquad \theta_1 =\frac{\sqrt{2} \zeta_1}{L}, \qquad
  \theta_2 =\frac{\sqrt{2} \zeta_2}{L}, \qquad
 \theta_3 =\frac{\sqrt{2} \zeta_3}{L},
\nonu
\eea
the Penrose limit of the  $AdS_4 \times Q^{1,1,1}$ becomes \cite{gns}
\bea
ds_{11}^2 & = & -4 dx^{+} dx^{-} +
 \sum_{i=1}^3 \left( dr^i dr^i -r^i r^i dx^{+}
dx^{+} \right) + \frac{1}{4}
\sum_{i=1}^3 \left( d\zeta_i^2 +\zeta_i^2 d\phi_i^2
- 2 \zeta_i^2 d\phi_i dx^{+} \right)
\nonu \\
&= &  -4 dx^{+} dx^{-} +\sum_{i=1}^3 \left( dr^i dr^i -r^i r^i dx^{+}
dx^{+} \right) + \frac{1}{4}
\sum_{i=1}^3 \left( d z_i d \bar{z}_i
+  i  \left( \bar{z}_i d z_i - z_i d \bar{z}_i \right) dx^{+} \right)
\label{ppwave}
\eea
where in the last line we introduce the complex coordinates $z_i = \zeta_i 
e^{i \phi_i}$. Since the metric has a covariantly constant null Killing
vector $\pa / \pa_{x^{-}}$, it is also pp-wave metric.
The pp-wave has a decomposition of the 
$\R^9$ transverse space into $\R^3 \times \R^2_1 \times \R^2_2 
\times \R^2_3$ where
$\R^3$ is parametrized by $r^i$ and $\R^2_1 \times \R^2_2 
\times \R^2_3 $ by $z_i$.
The symmetries of this background are the $SO(3)$ rotations in $\R^3$ and
a $U(1)_1 \times U(1)_2 \times U(1)_3$ symmetry corresponding to 
the rotations of
$\R^2 \times \R^2 \times \R^2$. In the gauge theory side,  
the $SO(3)$ symmetry corresponds to the subgroup of
the $SO(2,3)$ conformal group and $U(1)_1 \times U(1)_2 \times U(1)_3$  
rotation 
charges $J_1, J_2$ and $J_3$ correspond to the linear combination of
$Q_1$ and $R$, $Q_2$ and $R$, and $Q_3$ and $R$
respectively.  $R$ is the $U(1)_R$ charge of the gauge theory and $Q_1,
Q_2$ and $Q_3$ are the Cartan generators of the $SU(2)_1 \times SU(2)_2 \times
SU(2)_3$ global symmetry of the dual superconformal field theory.  
Note that the pp-wave geometry (\ref{ppwave}) in the scaling limit produces to
the ${\cal N}=8$ maximally supersymmetric pp-wave solution of
$AdS_4 \times \S^7$ \cite{kow,op,fk} through
$z_i = e^{i x^{+}} w_i$ and $\bar{z}_i = e^{-i x^{+}} \bar{w}_i$ \cite{go}
\bea
ds_{11}^2 = -4 d x^{+} d x^{-} +\sum_{i=1}^9  dr^i dr^i -
\left( \sum_{i=1}^3 r^i r^i  + \frac{1}{4}  \sum_{i=4}^9 
r^i r^i \right) dx^{+} dx^{+} 
\nonu
\eea
after an 
 $U(1)_1 \times U(1)_2 \times U(1)_3$ rotation in the $\R^2 \times \R^2 
\times \R^2$ plane \cite{gns}. 
The supersymmetry enhancement in the Penrose limit
implies that a hidden ${\cal N}=8$ supersymmetry is present in the 
corresponding subsector of the dual ${\cal N}=2$ superconformal field 
theory.  In the next section, we provide a description
of how to understand the excited KK states in the supergravity theory that 
corresponds in the dual superconformal field theory to operators with a
given conformal dimension. 

\section{Gauge Theory Spectrum}

The 11-dimensional supergravity theory in $ AdS_4 \times Q^{1,1,1}$
is dual to the ${\cal N}=2$ gauge theory encoded in a quiver diagram
with gauge group
$SU(N)_1 \times SU(N)_2 \times SU(N)_3$ 
with three series of chiral fields $A_i, i=1,2$ transforming in the 
$\bf (N, \overline{N},1)$ color representation, 
$B_j, j=1,2$ transforming in the 
$\bf (1,N, \overline{N})$ color representation and $C_k, k=1,2$
transforming in the $\bf (\overline{N},1,N)$ color representation 
\cite{fabbrietal}.
At the fixed point, these chiral superfields have conformal weight $1/3$ and
transform as $({\bf 2},{\bf 1},{\bf 1}), ({\bf 1},{\bf 2},{\bf 1})$ 
and $({\bf 1},{\bf 1},{\bf 2})$ under the $SU(2)_1 \times SU(2)_2 
\times SU(2)_3$ global symmetry.
Chiral fields $A_1$ and  $A_2$ are doublets of $SU(2)_1$ global symmetry,
 $B_1$ and  $B_2$ are doublets of $SU(2)_2$ global symmetry,
and $C_1$ and  $C_2$ are doublets of $SU(2)_3$ global symmetry.
There exist three kinds of supermultiplets in an ${\cal N}=2$ $AdS_4$
superspace: massless multiplets, short multiplets and long multiplets 
\cite{fabbri1}.
We identify states in the supergravity with operators in the gauge theory
and focus only on the bosonic excitations of the theory.  
In each multiplet, we specify 
field contents, a representation with $U(1)$ charge, conformal
weight and $U(1)_R$ $R$-charge.

$\bullet$ {\bf Massless multiplets} \cite{fabbri1,mer}

$1)$ Massless graviton multiplet: \footnote{Here we denote the field 
contents of each multiplet by their spin \cite{ffgt}. For massless 
graviton multiplet
$(2,\frac{3}{2},\frac{3}{2},1)$, there 
are one spin 2, two spin $3/2$ and one spin 1.}
$(2,\frac{3}{2},\frac{3}{2},1), \qquad
({\bf 1}, {\bf 1}, {\bf 1})_{\bf 0}, \qquad
 \Delta=2, \qquad R=0.
$ 

There exists a stress-energy tensor superfield $T_{\al \be}(x, \theta)$ 
that satisfies
the equation for conserved current $D_{\al}^{+} T^{\al \be}(x, \theta) = 
D_{\al}^{-}
T^{\al \be}(x, \theta)=0$. \footnote{
For three dimensional ${\cal N}=2$ 
superfields, we define $\theta_{\al}^{\pm}=
\frac{1}{\sqrt{2}} \left( \theta_{\al}^1 \pm i \theta_{\al}^2 \right)$
and the superderivatives are defined as
$D^{\pm \al} = \frac{\pa}{\pa \theta_{\al}^{\pm}} +\frac{1}{2}
(\gamma^m)^{\al}_{\be} \theta^{\be \pm} \pa_{m}$.} 
This $T_{\al \be}(x, \theta)$ 
is a singlet with respect to 
the flavor group $SU(2)_1 \times SU(2)_2 
\times SU(2)_3$ and its conformal dimension is 2. Moreover $R$ charge is 0.
So this corresponds to the massless graviton multiplet  that 
propagates in the $AdS_4$ bulk \cite{fabbrietal}. 

$2)$ Massless vector multiplet:
$(1,1,1,0,0), \qquad
({\bf 3}, {\bf 1}, {\bf 1})_{\bf 0}, \qquad
 \Delta=1, \qquad R=0.
$
 
There exists a conserved vector current, a scalar superfield 
$J_{SU(2)_i}(x, \theta)$,
to each generator of the flavor
symmetry group  through Noether theorem satisfying the conservation
equations $D^{+ \al} D^{+}_{ \al} J_{SU(2)_i}(x, \theta) = 
D^{- \al} D^{-}_{ \al}  J_{SU(2)_i}(x, \theta)= 0 $. 
This $J_{SU(2)_1}(x, \theta)$ transforms  in the adjoint  representation
of the first factor $SU(2)_1$ of 
the flavor group and its conformal dimension is 1 with vanishing
$R$-charge.
This corresponds to the massless vector multiplet  
propagating in the $AdS_4$ bulk \cite{fabbrietal}. 
There are also two massless vector multiplets denoted by
$({\bf 1}, {\bf 3}, {\bf 1})_{\bf 0}$ and 
$ ({\bf 1}, {\bf 1}, {\bf 3})_{\bf 0}$
coming from the permutation of three $SU(2)$ flavor groups.

$3)$ Two massless vector multiplets:
$(1,1,1,0,0), \qquad
({\bf 1}, {\bf 1}, {\bf 1})_{\bf 0}, \qquad
 \Delta=1, \qquad R=0.
$
 
It is known that two Betti currents of $Q^{1,1,1}$ are written down from
the toric description \cite{fabbrietal} 
and they are conserved.  Their conformal
dimension is 1 and $R$-charge is zero. Therefore Betti currents correspond to
additional massless vector multiplets. Therefore massless multiplets 1), 2)
and 3) saturate
the unitary bound and have a conformal weight(or dimension) related to the 
maximal spin. 

$\bullet$ {\bf Short multiplets} \cite{fabbri1,mer}

It is known that 
the dimension of the scalar operator in terms of energy labels,
in the dual SCFT corresponding
$AdS_4 \times Q^{1,1,1}$ is
\bea
\Delta = \frac{3}{2} + \frac{1}{2} \sqrt{1 + \frac{m^2}{4}} =
\frac{3}{2} + \frac{1}{2} \sqrt{45 + \frac{E}{4} -6 \sqrt{36 +E}} .
\label{delta}
\eea 
The first equation in (\ref{delta}) 
comes from the relation between the lowest energy
eigenvalue and the mass which appears in the $AdS_4$ wave equation.
The relation of d'Alembertian in $AdS_4$
to Casimir operator was obtained in \cite{nicolai,cfh}. 
The second relation in (\ref{delta}) comes from the formula of mass 
$m^2= E +176 -24 \sqrt{E +36}$ in \cite{df} where 
the normalization for this is
$(\Box -32 +m^2) S =0$ for scalar field $S$.
The energy spectrum on $Q^{1, 1, 1}$ exhibits an interesting feature
which is relevant to superconformal algebra and it is given by
\cite{pope,ahnplb}
\bea
E =  32 
\left( \sum_{i=1}^3 l_i(l_i+1)-s^2 \right)
\label{energy}
\eea
where the eigenvalue $E$ is classified by
$U(1)$ charge $s$ and spin(or angular momentum) 
$l_i$'s under $SU(2)_1 \times SU(2)_2 
\times SU(2)_3$ and the eigenmodes occur in $({\bf 2l_1+1},
{\bf 2l_2+1},{\bf 2l_3+1})_{\bf s}$ dimensional representation and  
$l_i \geq |s|, s=0, \pm \frac{1}{2}, \pm 1, \cdots$.
The $U(1)$ part of the isometry goup of $Q^{1, 1, 1}$
acts by shifting $U(1)$ charge $s$. 
The integer $R$-charge, $R$ is related to $U(1)$ charge $s$
by $s=R/2$. 
Let us take $R \geq 0$.  One can find the lowest value of $\Delta$ is
$R$ corresponding to a mode scalar with $l_i=s$ because $E$ 
becomes 
$32(2 s^2 +3s )$ and by plugging back to (\ref{delta})
then one obtains $\De=R$. 
Thus we find
a set of operators filling out a ${\bf (R+1, R+1, R+1)_{\frac{R}{2}}}$ 
multiplet of
$SU(2)_1 \times SU(2)_2 \times SU(2)_3 \times U(1)$ where a 
subscript is $U(1)$ charge
$R/2$ and the number ${\bf R+1}$ is 
the dimension of each $SU(2)$ representation.
The condition 
$\Delta=R$ saturates the bound on $\Delta$ from superconformal algebra.
The fact that the $R$-charge of a chiral operator is equal to
the conformal 
dimension was observed in \cite{bhk} in the context of $R$ symmetry gauge
field. 

$1)$ One hypermultiplet: $(\frac{1}{2},\frac{1}{2},0,0,0,0)$
\bea
({\bf R+1}, {\bf R+1}, {\bf R+1})_{\bf \frac{R}{2}}, \qquad
 \Delta=R.
\nonu
\eea 

It was shown in \cite{fabbrietal} that 
from the harmonic analysis on $Q^{1,1,1}$ and the spectrum of 
$SU(2)_1 \times SU(2)_2 \times SU(2)_3$ representation of the $OSp(2|4)$ 
hypermultiplets, the hypermultiplet of conformal dimension $\De=R$ and
$U(1)$ charge $s=R/2$ should be in the representation $l_i=s=R/2$. 
According to \cite{fabbri1}, 
the information on the Laplacian eigenvalues allows us 
to get the spectrum of hypermultiplets of the theory corresponding to
the chiral operators of the SCFT.
This part of spectrum was given in \cite{fabbrietal} and the form of operators
is 
\bea
\mbox{Tr} \Phi_{\mbox{c}} \equiv \mbox{Tr} (ABC)^R
\label{chiral}
\eea
where the flavor $SU(2)$'s indices 
are totally symmetrized and the chiral superfield
$\Phi_{\mbox{c}}(x, \theta)$ satifies $D_{\al}^{+} 
\Phi_{\mbox{c}}(x, \theta) =0$. 
The hypermultiplet spectrum in the KK harmonic expansions on $Q^{1,1,1}$
agrees with the chiral superfield predicted by the 
conformal gauge theory.
Based on the work of \cite{cfh} on a generic $OSp(2|4)$ representation, 
the operators with protected dimensions are 
related to shortenings and they are classified as three categories:
1) chiral superfields, 2) conserved currents and 
3) semi-conserved currents. The definitions of these fields 
are given in \cite{cddf}.
From this, the dimension of $ABC$ should be 1 to match the spectrum.
One realizes that the chiral fields $A_i, B_j, C_k$ have conformal
weight 1/3.

$2)$ One short graviton multiplet:
$(2, \frac{3}{2}, \frac{3}{2}, \frac{3}{2}, 1, 1, 1, \frac{1}{2})$ \cite{mer}
\bea
({\bf R+1}, {\bf R+1}, {\bf R+1})_{\bf \frac{R}{2}}, \qquad
 \Delta=R+2.
\nonu
\eea

The gauge theory interpretation of this multiplet is obtained by
adding a dimension 2 singlet operator with respect to 
flavor group into the above chiral superfield 
$ \Phi_{\mbox{c}}(x, \theta)$.
We consider
$
\mbox{Tr} \Phi_{\al \be} \equiv \mbox{Tr} \left( T_{\al \be}  \Phi_{\mbox{c}} 
\right),$
where $T_{\al \be}(x, \theta)$ 
is a stress energy tensor we have discussed before
and $ \Phi_{\mbox{c}}(x, \theta)$ is a chiral superfield in (\ref{chiral}). 
All color indices are symmetrized before taking the contraction.
This composite operator satisfies $D^{+}_{\al} 
\Phi^{\al \be}(x, \theta) =0 $.

$3)$ One short vector multiplet:
$(1, \frac{1}{2}, \frac{1}{2}, \frac{1}{2}, 0, 0, 0)$ \cite{mer}
\bea
({\bf R+3}, {\bf R+1}, {\bf R+1})_{\bf \frac{R}{2}}, \qquad
 \Delta=R+1.
\nonu
\eea
 
One can construct the following gauge theory object corresponding to
the short vector multiplet:
$
\mbox{Tr} \Phi  \equiv \mbox{Tr} \left( J_{SU(2)_1}  \Phi_{\mbox{c}} 
\right),$
where $J_{SU(2)_1}(x, \theta)$ 
is a conserved vector current transforming in the adjoint representation
of $SU(2)_1$ flavor group as before in massless vector 
multiplet and $ \Phi_{\mbox{c}}(x, \theta)$ is a chiral superfield 
in (\ref{chiral}). 
There exist
two other short multiplets by $({\bf R+1},{\bf R+3}, 
{\bf R+1})_{\bf
\frac{R}{2}}$ and $({\bf R+1},{\bf R+1}, {\bf R+3})_{\bf
\frac{R}{2}}$ due to the permutation of flavor groups. 
In this case, we have
$D^{+ \al} D^{+}_{\al} \Phi(x, \theta) =0$.  
There is another short vector multiplet described by $({\bf R+1},
{\bf R+1}, {\bf R+1})_{\bf \frac{R}{2}}, \Delta=R$. There 
exist three short gravitino multiplets$(\frac{3}{2},1,1,1,\frac{1}{2},
\frac{1}{2},\frac{1}{2},0)$ \cite{mer}
specified by 
\bea
({\bf R},
{\bf R+1}, {\bf R+1})_{\bf \frac{R}{2} +1}, \qquad
({\bf R+1},
{\bf R+2}, {\bf R+2})_{\bf \frac{R}{2} -1}, \qquad
({\bf R},
{\bf R+1}, {\bf R+1})_{\bf \frac{R}{2} +1}
\nonu
\eea
with conformal weight, $\Delta = R+3/2$. 
Therefore the short multiplets 1), 2) and 3) 
saturate the unitary bound and have a 
conformal dimension related to the $R$-charge and maximal spin. 
In superfield language, the $\theta$ expansion of the superfield
is shortened by imposing a suitable differential constraint, invariant
with respect to Poincare supersymmetry.   

$\bullet$ {\bf Long multiplets} \cite{mer}

Although the dimensions of nonchiral operators are in general irrational,
there exist special integer values of $k_i$ such that
for $l_i=k_i+R/2$, one can see the Diophantine like condition \cite{ahnplb}, 
\bea
-2(k_1 k_2+k_2 k_3+k_3 k_1)+\sum_{i=1}^3 (k_i^2- k_i)=0
\label{dequation}
\eea
make $\sqrt{36+E}$ be equal to
$4R+2(2\sum_{i=1}^3 k_i +3)$. 
This resembles the one in $AdS_5 \times T^{1,1}$ \cite{go}.
Furthermore in order to make 
the dimension be rational(their conformal dimensions are protected), 
$45 + E/4 -6 \sqrt{36 +E}$  in (\ref{delta}) should be square of 
something. It turns out this is the case without any further restrictions on
$k_i$'s. Therefore we  have $\De=R+\sum_i^3 k_i$ which is $\De_{+}$ for
$\De \geq 3/2$ and $\De_{-}$ for $\De \leq 3/2$. 
This is true if we are describing states with finite $\Delta$ and $R$.
Since we are studying the scaling limit $\Delta, R \rightarrow \infty$,
we have to modify the above analysis.
This constraint (\ref{dequation}) comes from the fact that the 
energy eigenvalue of the Laplacian on $Q^{1,1,1}$ for the supergravity mode
(\ref{energy})
takes the form 
\bea
E = \sum_{i=1}^3 \left( k_i^2 + (R+1)k_i \right) + 
\frac{R}{2} \left( R + 3 \right).    
\label{energyr}
\eea
One can show that the conformal weight of the long vector multiplet $A$
below becomes rational if the condition (\ref{dequation}) is satisfied.  
This multiplet has the following properties.

$1)$ One long vector multiplet $A$:
$(1,\frac{1}{2},\frac{1}{2},\frac{1}{2},\frac{1}{2},0,0,0,0,0)$
\bea
({\bf 2k_1 + R+1}, {\bf 2k_2 + R+1}, {\bf 2k_3 + R+1})_{\bf \frac{R}{2}}, 
\qquad
\Delta= -\frac{3}{2} +\frac{1}{4} \sqrt{E+36}.
\nonu 
\eea

However, as we take the limit of $R \rightarrow \infty$, 
this constraint (\ref{dequation}) is relaxed. The combination of 
$\Delta-R$ is given by   
\bea
\Delta-R = k_1 +k_2 +k_3 + {\cal O}(\frac{1}{R})
\label{lowest}
\eea
where the right hand side is definitely rational and they are integers.
So the constraint  (\ref{dequation}) is not relevant in the
subsector of the Hilbert space we are interested in. 
Candidates for such states in the gauge theory side are given in terms of
semi-conserved superfields \cite{fabbrietal}. 
Although they are not chiral primaries, their
conformal dimensions are protected. The ones we are interested in take the
following form,
\bea
\mbox{Tr} \Phi_{\mbox{s.c.}} \equiv
\mbox{Tr} \left[  \left( J_{SU(2)_1} \right)^{k_1} \left( J_{SU(2)_2} 
\right)^{k_2} 
\left( J_{SU(2)_3} \right)^{k_3}  \left(A B C \right)^R \right]
\label{semi}
\eea   
where the scalar superfields $J_{SU(2)_i}(x,\theta)$ transform 
in the adjoint representation of flavor group $SU(2)_i$ and satisfy
$D^{+ \al} D^{+}_{\al} J_{SU(2)_i}(x,\theta) =
D^{- \al} D^{-}_{\al} J_{SU(2)_i}(x,\theta)=0 $ with conformal dimension 1
and zero $U(1)_R$ charge. Also we have $D^{+ \al} D^{+}_{\al} \Phi{
\mbox{s.c.}}(x, \theta)=0$. 
Since the singleton superfields $A_{i, b}^{a}$
carry an index $a$ in the ${\bf N}$ of $SU(N)_1$ and an
index $b$ in the ${\bf \overline{N}}$ of the $SU(N)_2$,
the fields
$B_{j, c}^{b}$
carry an index $b$ in the ${\bf N}$ of $SU(N)_2$ and an
index $c$ in the ${\bf \overline{N}}$ of the $SU(N)_3$,
and
$C_{k, a}^{c}$
carry an index $c$ in the ${\bf N}$ of $SU(N)_3$ and an
index $a$ in the ${\bf \overline{N}}$ of the $SU(N)_1$,
one can construct the following conserved flavor
currents transforming
$({\bf 3},{\bf 1},{\bf 1})_{\bf 0}, ({\bf 1},{\bf 3},{\bf 1})_{\bf 0}$ 
and $ ({\bf 1},{\bf 1},{\bf 3})_{\bf 0}$ respectively 
\bea
 \left( J_{SU(2)_1} \right)_
{i_1}^{j_1} & = & A^{j_1} \overline{A}_{i_1} -\frac{\delta_
{i_1}^{j_1}}{2} A \overline{A},\nonu \\ 
 \left( J_{SU(2)_2} \right)_
{i_2}^{j_2} & = & B^{j_2} \overline{B}_{i_2} -\frac{\delta_
{i_2}^{j_2}}{2} B \overline{B}, \nonu \\
\left( J_{SU(2)_3} \right)_
{i_3}^{j_3} & = & C^{j_3} \overline{C}_{i_3} -\frac{\delta_
{i_3}^{j_3}}{2} C \overline{C},
\nonu 
\eea
where the color indices are contracted in the right hand side.
Note that the conformal dimension of these currents is not
the one of naive sum of $A$ and $\overline{A}$,
$B$ and $\overline{B}$ and 
$C$ and $\overline{C}$. 
As we discussed in the last section, supergravity theory in $AdS_4 \times
Q^{1,1,1}$ acquires an enhanced ${\cal N}=8$ superconformal symmetry in the
Penrose limit. This implies that the spectrum of the gauge theory 
operators in this subsector should fall into
${\cal N}=8$ multiplets. We expect that both the chiral primary
fields of the form  $\mbox{Tr} (ABC)^R$ and the semi-conserved multiplets
of the form (\ref{semi}) combine into make ${\cal N}=8$ multiplets in this 
limit. Note that for finite $R$, the semi-conserved multiplets
should obey the Diophantine constraint (\ref{dequation}) in order for them to
possess rational conformal weights. 

In the remaining multiplets we consider the following particular
representations in the global symmetry group with $U(1)$ charge:
$
({\bf 2k_1 + R+1}, {\bf 2k_2 + R+1}, {\bf 2k_3 + R+1})_{\bf \frac{R}{2}},
$

$2)$ One long vector multiplet $Z$:
$(1,\frac{1}{2},\frac{1}{2},\frac{1}{2},\frac{1}{2},0,0,0,0,0), \qquad
\Delta= \frac{1}{2} +\frac{1}{4} \sqrt{E+32R+4}
$

For finite $R$, in order to make the dimension of this multiplet
rational the expression
$E+32R+4$ should be square of something. By plugging the expression 
of $E$ (\ref{energyr}), 
it turns out $\sqrt{E+32R+4}=4R+4(k_1+k_2+k_3)+2$ with same 
constraint (\ref{dequation}). So the combination of $\Delta-R$
becomes 
\bea
\Delta-R = 1+ k_1 +k_2 +k_3 +{\cal O} \left(\frac{1}{R} \right).
\nonu
\eea
Since we do not have any singlet of conformal dimension 1  
with respect to the flavor group, one cannot increase a conformal dimension
by simply tensoring any extra superfields into a semi-conserved current
in order to match the spectrum.
So the only way to do this is to increase the number of 
conserved scalar superfield. 
In order to produce the following gauge theory operator 
$
 \mbox{Tr} \left(
J_{SU(2)_1} \Phi_{\mbox{s.c.}} \right)$
corresponding to
this vector multiplet, one can think of a 
higher dimensional representation in
the first factor $SU(2)_1$ in the global symmetry. 
Then the constraint coming
from the requirement of rationality of conformal dimension is also changed
for finite $\Delta$ and $R$.

$3)$ One long graviton multiplet $h$:$(2, \frac{3}{2},
\frac{3}{2},\frac{3}{2},\frac{3}{2},1,1,1,1,
1,1,\frac{1}{2},\frac{1}{2},\frac{1}{2},\frac{1}{2},0)$
\bea
\Delta= \frac{1}{2} +\frac{1}{4} \sqrt{E+36}.
\nonu
\eea

For finite $R$ with rational dimension, 
after inserting the $E$ into the above, 
we will arrive at the relation with same constraint (\ref{dequation})
which is greater than (\ref{lowest}) by 2:
\bea
\Delta-R = 2+ k_1 +k_2 +k_3 +{\cal O} \left(\frac{1}{R} \right).
\label{delta2}
\eea
The gauge theory interpretation of this multiplet is
quite simple. If we take  a semi-conserved current 
$\Phi_{\mbox{s.c.}}(x, \theta)$ defined in (\ref{semi})
and multiply it by a stress-energy tensor superfield 
$T_{\al \be}(x, \theta)$ that is a singlet with respect to the flavor group,
namely 
\bea
\mbox{Tr} \left( T_{\al \be}  \Phi_{\mbox{s.c.}} 
\right),
\nonu
\eea
we reproduce the right $OSp(2|4) \times SU(2)_1 \times SU(2)_2 \times SU(2)_3$
representations of the long vector multiplet.
Also one can expect that other candidate for this multiplet with different 
representation by multiplying a semi-conserved current into a quadratic
 conserved scalar superfield: 
$\mbox{Tr} \left( J_{SU(2)_i} J_{SU(2)_j}  
\Phi_{\mbox{s.c.}} \right)$. The constraint for finite $\Delta$ and $R$ is
shifted as $k_i \rightarrow k_i+1$ and $k_j \rightarrow k_j +1$.

$4)$ One long vector multiplet  $W$:
$(1,\frac{1}{2},\frac{1}{2},\frac{1}{2},\frac{1}{2},0,0,0,0,0), \qquad
\Delta= \frac{5}{2} +\frac{1}{4} \sqrt{E+36}
$.

In this case, we get $\Delta -R$ by adding 2 to the one in (\ref{delta2})
\bea
\Delta-R = 4+ k_1 +k_2 +k_3 +{\cal O} \left(\frac{1}{R} \right).
\nonu
\eea
One can describe corresponding gauge theory operator by taking
quardratic stress-energy tensor $ T_{\al \be} T^{\al \be}(x, \theta)$
and mutiplying it into a semi-conserved 
current $\Phi_{\mbox{s.c.}}(x, \theta)$
we have seen before in order to match with 
 the conformal dimension. That is, one obtains  
\bea
\mbox{Tr} \left( T_{\al \be} T^{\al \be}  
\Phi_{\mbox{s.c.}} \right).
\nonu
\eea
Similarly one can construct the following gauge theory operators
related to this vector multiplet 
$\mbox{Tr} \left(T_{\al \be} J_{SU(2)_i} J_{SU(2)_j}  
\Phi_{\mbox{s.c.}} \right)$ or $
\mbox{Tr} \left( J_{SU(2)_i} J_{SU(2)_j} 
 J_{SU(2)_k} J_{SU(2)_l}   
\Phi_{\mbox{s.c.}}\right)$. In the former, one can see the shift of
$k_{i,j} \rightarrow k_{i,j} +1$ and for the latter, the shift of
$k_{i,j,k,l} \rightarrow k_{i,j,k,l} +1$. 

$5)$ One long vector multiplet $Z$:
$(1,\frac{1}{2},\frac{1}{2},\frac{1}{2},\frac{1}{2},0,0,0,0,0),
\qquad \Delta= \frac{1}{2} +\frac{1}{4} \sqrt{E+4}
$

Although there exists no rational dimension for this case with any choice of
$k_i$'s when $\Delta$ and $R$ are finite, 
the combination of $\Delta -R$ with Penrose limit 
$R \rightarrow \infty$ in the gauge theory side becomes
\bea
\Delta-R = 2+ k_1 +k_2 +k_3 +{\cal O} \left(\frac{1}{R} \right).
\nonu
\eea  
In addition to the above 1), 2), 3), 4) and 5) multiplets,
there are also 
two long gravitino multiplets 
\cite{mer} $\chi^{+}$$(\frac{3}{2},1,1,1,1,
 \frac{1}{2}, \frac{1}{2}, \frac{1}{2}, \frac{1}{2}, \frac{1}{2},
\frac{1}{2},0,0,0,0)$
characterized by $
\Delta  =  -\frac{1}{2} +\frac{1}{4} \sqrt{E \pm 16R+32}
$
and  two long gravitino multiplets $\chi^{-}$ characterized by 
$
\Delta  =  \frac{3}{2} +\frac{1}{4} \sqrt{E \pm 16R+32}
$. Similar anaylsis can be done in this case. Although for finite 
$\Delta$ and $R$, both do not provide rational conformal dimensions,
in the Penrose limit there is no constraint on the integer values and
$R \rightarrow \infty$ limit will give us a rational conformal dimension. 

\section{Conclusion}

We described an explicit example of an ${\cal N}=2$
three-dimensional 
superconformal field theory that has a subsector of the Hilbert space
with enhanced ${\cal N}=8$ superconformal symmetry, in the large $N$ limit.
This subsector of gauge theory is achieved by Penrose limit
which  constrains strictly the states of the gauge theory to those
whose conformal dimension and $R$ charge diverge in the large $N$ limit
but possess finite value $\Delta-R$.
We predicted for the spectrum of $\Delta-R$ of the ${\cal N}=2$
superconformal field theory and proposed how the exicited states in the
supergravity correspond to
gauge theory operators. In particular, both the chiral multiplets 
(\ref{chiral}) and
semi-conserved multiplets (\ref{semi}) of ${\cal N}=2$ supersymmetry should
combine into ${\cal N}=8$ chiral multiplets. 
It would be interesting to study other types of compatification on M-theory
in the Penrose limit which will provide pp-wave background and see what the 
corresponding gauge theory operators possessing a hidden ${\cal N}=8$
supersymmetry are.

\vskip2cm
$\bf Acknowledgements$

This research was supported 
by 
grant 2000-1-11200-001-3 from the Basic Research Program of the Korea
Science $\&$ Engineering Foundation.


\begin{thebibliography}{[00]}
\bibitem{bmn}  D. Berenstein, J. Maldacena and H. Nastase, JHEP {\bf 0204}
(2002) 013, {\tt hep-th/0202021}.
\bibitem{bfhpetal}  M. Blau, J. Figueroa-O'Farrill, C.M. Hull
 and G. Papadopoulos, 
{\tt hep-th/0201081}.
\bibitem{bfhpetal1}  M. Blau, J. Figueroa-O'Farrill, C.M. Hull
 and G. Papadopoulos, JHEP {\bf 0201} (2001) 047,
{ \tt hep-th/0110242}.
\bibitem{mt} R.R. Metsaev and A.A. Tseytlin, {\tt hep-th/0202109}.
\bibitem{betal} M. Blau, J. Figueroa-O'Farrill and G. Papadopoulos, 
{\tt hep-th/0202111}.
\bibitem{bhk} D. Berenstein, C.P. Herzog and I.R. Klebanov, 
{\tt hep-th/0202150}.
\bibitem{ikm} N. Itzhaki, I.R. Klebanov and S. Mukhi, 
JHEP {\bf 0203} (2002) 048,
{\tt hep-th/0202153}.
\bibitem{go} J. Gomis and H. Ooguri, {\tt hep-th/0202157}.
\bibitem{rt}J.G. Russo and A.A. Tseytlin, 
JHEP {\bf 0204} (2002) 021,
{\tt hep-th/0202179}.
\bibitem{zs} L.A. Pando Zayas and J. Sonnenschein, {\tt hep-th/0202186}.
\bibitem{hks} M. Hatsuda, K. Kamimura and M. Sakaguchi, {\tt
hep-th/0202190}.
\bibitem{as} M. Alishahiha and M.M. Sheikh-Jabbari, {\tt hep-th/0203018}.
\bibitem{bp} M. Billo and I. Pesando, {\tt hep-th/0203028}.
\bibitem{kprt} N. Kim, A. Pankiewicz, S.-J. Rey and S. Theisen, 
{\tt hep-th/0203080}.
\bibitem{clp} M. Cvetic, H. Lu and C.N. Pope, {\tt hep-th/0203082}. 
\bibitem{tt} T. Takayanagi and S. Terashima, {\tt hep-th/0203093}.
\bibitem{gns} U. Gursoy, C. Nunez and M. Schvellinger, {\tt hep-th/0203124}.
\bibitem{fk} E. Floratos and A. Kehagias, {\tt hep-th/0203134}.
\bibitem{m} J. Michelson, {\tt hep-th/0203140}.
\bibitem{g} R. Gueven, {\tt hep-th/0203153}.
\bibitem{dgr} S.R. Das, C. Gomez and S.-J. Rey, {\tt hep-th/0203164}.
\bibitem{ch} C.-S. Chu and P.-M. Ho, {\tt hep-th/0203186}.
\bibitem{clp1}  M. Cvetic, H. Lu and C.N. Pope, {\tt hep-th/0203229}.
\bibitem{dp} A. Dabholkar and S. Parvizi, {\tt hep-th/0203231}.
\bibitem{bgnn} D. Berenstein, E. Gava, J. Maldacena, K.S. Narain and
H. Nastase, {\tt hep-th/0203249}.
\bibitem{gh} J.P. Gauntlett and C.M. Hull, {\tt hep-th/0203255}.
\bibitem{lp} P. Lee and J. Park, {\tt hep-th/0203257}. 
\bibitem{lv} H. Lu and J.F. Vazquez-Poritz, {\tt hep-th/0204001}.
\bibitem{hks1}  M. Hatsuda, K. Kamimura and M. Sakaguchi, 
{\tt hep-th/0204002}.
\bibitem{kp} E. Kiritsis and B. Pioline, {\tt hep-th/0204004}.
\bibitem{kn} A. Kumar, R.R. Nayak and Sanjay, {\tt hep-th/0204025}.
\bibitem{lor} R.G. Leigh, K. Okuyama and M. Rozali, {\tt hep-th/0204026}. 
\bibitem{b} D. Bak, {\tt hep-th/0204033}.
\bibitem{gkp} S.S. Gubser, I.R. Klebanov and
A.M. Polyakov, {\tt hep-th/0204051}.
\bibitem{st} K. Skenderis and M. Taylor, {\tt hep-th/0204054}.
\bibitem{sv} M. Spradlin and A. Volovich, {\tt hep-th/0204146}.
\bibitem{mrv} S. Mukhi, M. Rangamani and E. Verlinde, {\tt hep-th/0204147}.
\bibitem{as1} M. Alishahiha and M.M. Sheikh-Jabbari, {\tt hep-th/0204174}.
\bibitem{bhln} V. Balasubramanian, M. Huang, T.S. Levi and
A. Naqvi, {\tt hep-th/0204196}.
\bibitem{ima} Y. Imamura, {\tt hep-th/0204200}. 
\bibitem{ft} S. Frolov and A.A. Tseytlin, {\tt hep-th/0204226}.
\bibitem{tt1} H. Takayanagi and T. Takayanagi, {\tt hep-th/0204234}.
\bibitem{dfv} R. D'Auria, P. Fre and P. van Nieuwenhuizen, Phys.Lett. 
{\bf B136} (1984) 347.
\bibitem{mer} P. Merlatti, 
Class.Quant.Grav. {\bf 18} (2001) 2797,
{\tt hep-th/0012159}.
\bibitem{fabbrietal} D. Fabbri, P. Fre, L. Gualtieri, C. Reina,
A. Tomasiello, A. Zaffaroni and A. Zampa, 
Nucl.Phys. {\bf B577} (2000) 547,
{\tt hep-th/9907219}.
\bibitem{ffgt} D. Fabbri, P. Fre, L. Gualtieri and P. Termonia,
Class.Quant.Grav. {\bf 17} (2000) 55,
{\tt hep-th/9905134}.
\bibitem{np} B.E.W. Nilsson and C.N. Pope, Class.Quant.Grav. {\bf 1}
(1984) 499.
\bibitem{pp} D. Page and C.N. Pope, Phys.Lett. {\bf B144} (1984) 346.
\bibitem{pp1} D. Page and C.N. Pope, Phys.Lett. {\bf B145} (1984) 337.
\bibitem{ahnplb} C. Ahn, Phys.Lett. {\bf B466} (1999) 171, 
{\tt hep-th/9908162}.
\bibitem{kow} J. Kowalski-Glikman, Phys.Lett. {\bf B134} (1984) 194.
\bibitem{op} J. Figueroa-O'Farrill and G. Papadopoulos, 
JHEP {\bf 0108 } (2001) 036,
{\tt hep-th/0105308}.
\bibitem{fabbri1}  D. Fabbri, P. Fre, L. Gualtieri and P. Termonia,
Nucl.Phys. {\bf B560} (1999) 617, 
{\tt hep-th/9903036}.
\bibitem{nicolai} H. Nicolai, Representations of Supersymmetry in Anti-De 
Sitter Space, Trieste School 1984:368 (QC178:S81:1984).
\bibitem{cfh} A. Ceresole, P. Fre and H. Nicolai, Class.Quant.Grav. {
\bf 2} (1985) 133. 
\bibitem{df} R. D'Auria and P. Fre, Annals Phys. {\bf 162} (1985) 372.
\bibitem{pope} C.N. Pope, Class.Quant.Grav. {\bf 1} (1984) L91.
\bibitem{cddf} A. Ceresole, G. Dall'Agata, R. D'Auria and S. Ferrara,
JHEP {\bf 0003} (2000) 011, {\tt hep-th/9912107}.
\end{thebibliography}
\end{document}